\documentclass[11pt]{article}

\usepackage{amsmath,amsthm,epsfig,epsf,psfrag,natbib}
\usepackage{graphicx,amssymb}
\usepackage{enumerate}
\usepackage[latin1]{inputenc}

\usepackage{amsbsy,lscape,epstopdf}
\usepackage{color,latexsym,amsfonts,bbm,amscd}

\usepackage{url,mathtools}

\usepackage{amsmath}
\usepackage{epsfig}
\usepackage{epsf}
\usepackage{psfrag}
\usepackage{natbib}
\usepackage{amssymb}
\usepackage[latin1]{inputenc}
\usepackage{rotating}
\usepackage{amsbsy}
\usepackage{lscape}
\usepackage{color}
\usepackage{amscd}
\usepackage{graphicx}
\usepackage{epstopdf}
\usepackage{hyperref}

\DeclarePairedDelimiter\ceil{\lceil}{\rceil}

\newtheorem{theorem}{Theorem}
\newtheorem{lemma}{Lemma}

\newtheorem{prop}{Proposition}

\newcommand{\be}{\begin{equation}}
\newcommand{\ee}{\end{equation}}

\newcommand{\bTheta}{\mbox{\boldmath $\theta$}}
\newcommand{\bas}{\begin{eqnarray*}}
\newcommand{\eas}{\end{eqnarray*}}

\newcommand{\ba}{\mbox{\bf a}}
\newcommand{\bb}{\mbox{\bf b}}

\newcommand{\tT}{{ \mathrm{\scriptscriptstyle T} }}

\begin{document}

\date{}
\title{\bf  Maximum multinomial likelihood estimation in compound mixture model
with application to malaria study}

 \author{Zhaoyang Tian,  \ Kun Liang,  \ Pengfei Li
 \footnote{Corresponding author. E-mail: pengfei.li@uwaterloo.ca.}
\\[1ex]
Department of Statistics and Actuarial Science\\
University of Waterloo,
Waterloo, ON, Canada, N2L 3G1\\
}
\maketitle

\abstract{
Malaria can be diagnosed by the presence of parasites and symptoms (usually fever) due to the parasites. In endemic areas, however, an individual may have fever attributable either to malaria or to other causes. Thus, the parasite level of an individual with fever follows a two-component mixture, with the two components corresponding to malaria and nonmalaria individuals. Furthermore, the parasite levels of nonmalaria individuals can be characterized as a mixture of a zero component and a positive distribution. In this article, we propose a nonparametric maximum multinomial likelihood approach for estimating the proportion of malaria using parasite-level data from two groups of individuals collected in two different seasons. We develop an EM-algorithm to numerically calculate the proposed estimates and further establish their convergence rates.  Simulation results show that the proposed estimators are more efficient than existing nonparametric estimators. The proposed method is used to analyze a malaria survey data.
\vskip0.5cm \noindent\textbf{Key words:} \textit{
Binomial likelihood; Clinical malaria; EM-algorithm; Empirical likelihood; Parasite level.
}

\vskip0.5cm\noindent\textbf{AMS 2000 Subject Classifications:}
\textit{Primary} \textbf{62G05}, \textit{Secondary} \textbf{62G20}.

\vskip0.5cm\noindent\textbf{Short title: }\textit{Nonparametric estimation in compound mixture model}
%

\section{ Introduction}
Despite the major progress in the fight against malaria, 
it remains an acute public health problem, 
particularly in sub-Saharan Africa. 
Millions of people are at risk of malaria each year throughout the world \citep{Bhatt15}. 
Malaria can be diagnosed by the presence of parasites and symptoms (usually fever) due to the parasites.
However, malaria is not the only disease that is associated with fever. 
Further, in areas of high endemicity, asymptomatic parasitaemia is very common, 
and it should not be identified as clinical malaria. 
Thus, it is challenging to develop accurate diagnosis methods for malaria.

The data from a cross-sectional survey of parasitemia and fever for children less than a year old in a village in the Kilombero district of Tanzania \citep{Kitua96}  is an excellent illustrating and motivating example. 
In one study of the survey, parasite levels in blood were collected in two seasons: dry season and wet season.
The prevalence of malaria varies between these seasons. 
In the dry season, the prevalence is very low, and the parasite levels collected can be considered to come from nonmalaria individuals. 
In the wet season, the prevalence is high due to the presence of mosquitoes. 
However, children can tolerate parasites without symptoms and 
may have fever from other causes. 
Hence,
the parasite levels collected in the wet season 
can be viewed as a mixture of those from people infected with malaria and those from nonmalaria individuals.
One problem of interest is to estimate the malaria frequency in the wet season based on the parasite levels collected from both the wet and dry seasons \citep{Smith1994}.

A unique feature of the data is that the parasite levels of some nonmalaria individuals are exactly zero. 
The parasite levels from individuals in the dry season (nonmalaria individuals) therefore follow a mixture of zero and a positive  distribution, and 
the parasite levels from individuals in the wet season
follow a compound mixture \citep{Qin05}. 
Specifically, let $X_1,\ldots,X_m$ be the parasite levels from the nonmalaria individuals (i.e.~individuals from the dry season)
and $Y_1,\ldots,Y_n$ be the parasite levels from the mixture of malaria and nonmalaria individuals (i.e.~individuals from the wet season). 
Then we have 
\begin{align}
X_1,...,X_m&\sim F_X(x)=pI(x\geq0)+(1-p)F_1(x),
\label{model1}\\
Y_1,...,Y_n&\sim F_Y(y)=(1-\lambda)\left\{pI(y\geq0)+(1-p)F_1(y)\right\}+\lambda F_2(y),
\label{model2}
\end{align}
where  $p$ is the proportion of nonmalaria individuals with zero parasite levels, 
$\lambda$ is the proportion of malaria individuals in the wet season, and
$F_1$ and $F_2$ are the cumulative distribution functions (cdfs) of positive parasite levels for the nonmalaria and malaria populations, respectively. 
We wish to estimate $(\lambda,p,F_1,F_2)$ without making additional assumptions on $F_1$ and $F_2$. 
There are at least two important applications for the estimate of $(\lambda,p,F_1,F_2)$. 
First, the estimate of $\lambda$, the estimated malaria frequency in the mixture sample, 
will be helpful  for the implementation of intervention programmes \citep{Vounatsou98}. 
Second, the estimator of $(\lambda,p,F_1,F_2)$ can be used to construct 
an estimator of the posterior probability that an individual in the mixture sample is infected with malaria
given his/her positive parasite level. 
See Section \ref{example.malaria} for more details. 

The compound structure seen in (\ref{model1}) and (\ref{model2}) does not appear only in the malaria study. 
It also appears in biomedical research and diagnostic practice, especially when there is no gold standard for the true positives.
See \citet{Smith1994} and \citet{Qin2017} for more examples.  
Using  (\ref{model1}) and (\ref{model2}), 
\citet{Smith1994} consider a model-based approach 
in which the relationship between the parasite level and 
malaria status (malaria or nonmalaria) is modelled 
through a logistic regression.
\citet{Qin05} observe that this relationship is equivalent 
to a density ratio model assumption on the probability density functions (pdfs) of $F_1$ and $F_2$. 
They further propose a semiparametric likelihood method
to estimate $(\lambda,p)$ under the density ratio model assumption. 
\citet{Vounatsou98} consider the estimation of $(\lambda,p)$
under a setup where the pdf ratio of $F_2$ and $F_1$ is a monotonically increasing function of the parasite level. 
They suggest first grouping the positive parasite levels 
into several ordered categories
and then estimating the unknown parameters by a Bayesian method. 
In summary,   the existing methods rely on
certain model assumptions for the pdfs of $F_1$ and $F_2$, 
which may not be robust to model misspecification. 
Further,  \citet{Vounatsou98}'s method requires ad-hoc grouping of 
the positive parasite levels. 

In this paper, we concentrate on the situation where there is no assumption on $F_1$ and $F_2$ except that they are cdfs. 
Under this setup,  
$(\lambda,p)$ can be naturally estimated by 
the binomial estimator \citep{Qin05}
as follows: 
\begin{align}
\label{binest1.malaria}
\tilde{p}&=\frac{m_0}{m},~\tilde{\lambda}=1-\frac{n_0}{n\tilde{p}},
\end{align}
where $m_0$ and $n_0$ are  the number of zeros  in  $\{X_1,\ldots,X_m\}$ and $\{Y_1,\ldots,Y_n\}$, respectively.
With the estimator of $(\lambda,p)$, 
$(F_1, F_2)$ can then be 
naturally estimated by 
\begin{align}
\label{binest2.malaria}
\tilde{F}_1(x)&=\tilde F_{X+}(x),~~
\tilde{F}_2^*(x)=(\tilde{\lambda}^*)^{-1}\left\{\tilde F_{Y+}(x) -(1-\tilde{\lambda}^*) \tilde{F}_1(x)\right\},
\end{align}
where 
$\tilde{\lambda}^*={\tilde{\lambda}}/\{1- \tilde{p}(1-\tilde{\lambda})\}$ and 
$
\tilde F_{X+}(x)
$ and $
\tilde F_{Y+}(x)
$
are the empirical cdfs of the positive values in  $\{X_1,\ldots,X_m\}$
and $\{Y_1,\ldots,Y_n\}$, respectively. 
To ensure the monotonicity of the estimator of $F_2(x)$, 
 one can apply isotonic regression to $\tilde{F}_2^*$ to obtain an estimator $\tilde F_2(x)$ of $F_2$, 
 which is guaranteed to be an increasing function. 
 See Section \ref{simulation.malaria} for more details.  
 
The estimator $(\tilde \lambda,\tilde p, \tilde{F}_1,\tilde{F}_2)$ is quite natural and simple to implement. 
However, the estimator $(\tilde\lambda,\tilde p,\tilde F_1)$ does not use the compound structure in (\ref{model2}). 
Further, the estimator $(\tilde \lambda,\tilde p, \tilde{F}_1,\tilde{F}_2)$ 
does not have a likelihood interpretation and may not be fully efficient. 
There is room for improvement. 
Motivated by the idea of the binomial likelihood \citep{Qin14}, 
we propose a maximum multinomial likelihood approach to  
estimate $(\lambda,p,F_1,F_2)$ simultaneously. 
This method takes the zero-inflated and compound structure in (\ref{model1}) and (\ref{model2})
into account. 
The resulting maximum multinomial likelihood estimator is shown to be consistent. 
Its asymptotic properties are also inverstigated. 
Simulation studies show that the proposed estimator is more efficient than $(\tilde \lambda,\tilde p, \tilde{F}_1,\tilde{F}_2)$.

The rest of the paper is organized as follows.  
In Section \ref{review.malaria}, 
we discuss the application of the empirical likelihood method \citep{Owen01}
and show that it fails to produce a consistent estimator for $(\lambda,F_2)$. 
We further review the idea of the binomial likelihood and its recent development. 
In Section \ref{method.malaria}, we propose the maximum multinomial likelihood estimator
for  $(\lambda,p,F_1,F_2)$ and establish its asymptotic properties. 
We also develop an EM algorithm to numerically calculate the estimate. 
In Section \ref{simulation.malaria}, we use simulation studies to compare the proposed maximum multinomial likelihood estimator
with $(\tilde \lambda,\tilde p, \tilde{F}_1,\tilde{F}_2)$ and demonstrate its superiority. 
Section \ref{example.malaria} applies our method to the real dataset from   \citet{Kitua96},
and  Section \ref{conclusion.malaria} provides concluding remarks.
For convenience of presentation, 
all the technical details are given in the Appendix.

\section{Empirical likelihood and maximum binomial likelihood}
\label{review.malaria}
\subsection{Failure of empirical likelihood method}
As a nonparametric likelihood method, the empirical likelihood method \citep{Owen01} seems 
a natural choice to estimate $(\lambda,p,F_1,F_2)$. 
However, we will show that it fails to produce a consistent estimator for $(\lambda,F_2)$. 

Let $m_+$ and $n_+$ be the numbers of positive values in  $\{X_1,\ldots,X_m\}$ and $\{Y_1,\ldots,Y_n\}$, respectively. 
Without loss of generality, we use $X_{1},\ldots,X_{m_+}$ to denote positive values in $\{X_1,\ldots,X_m\}$
and $Y_1,\ldots,Y_{n+}$ to denote positive values in $\{Y_1,\ldots,Y_n\}$. 
For illustration, we consider the case where there is no tie in $\{X_1,\ldots,X_{m_+}, Y_1,\ldots,Y_{n_+}\}$.  

Denote $a_i=dF_1(X_i)$ for $i=1,\ldots,m_+$ and $a_{j+m_+}=dF_1(Y_j)$ and $b_{j}=dF_2(Y_j)$ for $j=1,\ldots,n_+$.
Let $\ba=(a_1,\ldots,a_{m_++n_+})^\tT$ and $\bb=(b_1,\ldots,b_{n_+})^{\tT}$.
Following the empirical likelihood principle \citep{Owen01},
the  log empirical likelihood of $(\lambda,p,\ba,\bb)$ is
$$
\tilde l_{el}(\lambda,p,\mathbf{a},\mathbf{b})
=m_0\log p
+\sum_{i=1}^{m_+} \log\left\{(1-p)a_i\right\}+n_0\log\left\{(1-\lambda)p\right\}
+\sum_{j=1}^{n_+}\log\left\{(1-\lambda)(1-p)a_{j+m_+}+\lambda b_j\right\}.
$$

The maximum empirical likelihood estimator of $(\lambda,p,\ba,\bb)$
is 
$$
(\tilde \lambda_{el},\tilde p_{el},\tilde \ba_{el},\tilde \bb_{el})=\arg\max_{\lambda,~p,~\mathbf{a},~\mathbf{b}} \tilde l_{el}(\lambda,p,\ba,\bb)
$$
subject to the constraints 
$$
\lambda\in[0,1],~~p\in[0,1],~~a_i\geq 0,~~b_j\geq 0,~~\sum_{i=1}^{m_++n_+}a_i=1,~~\sum_{j=1}^{n_+}b_j=1.
$$
The maximum empirical likelihood estimators of $F_1(x)$ and $F_2(x)$
are
$$
\tilde F_{1,el}(x)=\sum_{i=1}^{m_+}\tilde a_{i,el}I(X_i\leq x)+\sum_{j=1}^{n_+}\tilde a_{j+m_+,el}I(Y_j\leq x),~~
\tilde F_{2,el}(x)=\sum_{j=1}^{n_+}\tilde b_{j,el}I(Y_j\leq x).
$$

The following proposition gives the closed form of $(\tilde \lambda_{el},\tilde p_{el},\tilde F_{1,el},\tilde F_{2,el})$. 
The proof is in Appendix A. 
\begin{prop}
\label{prop1}
Suppose there is no tie in the positive observations in $\{X_1,\ldots,X_m, Y_1,\ldots,Y_n\}$. 
Then we have  
$$
\tilde \lambda_{el}=\frac{n_+}{n}, ~~
\tilde p_{el}=\frac{m_0+n_0}{m+n_0},
$$
and 
$$
\tilde F_{1,el}(x)=\frac{1}{m_+}\sum_{i=1}^{m_+} I(X_i\leq x),~~
\tilde F_{2,el}(x)=\frac{1}{n_+}\sum_{j=1}^{n_+} I(Y_j\leq x).
$$
\end{prop}

Proposition \ref{prop1} implies  that 
$$
\tilde\lambda_{el}\to (1-\lambda)(1-p)+\lambda,~~
\tilde F_{2,el}(x) \to (1-\lambda^*)F_1(x)+\lambda^*F_2(x)
$$ 
in probability, where 
$\lambda^*={\lambda}/\{1- p(1-\lambda)\}$. 
Therefore, when $0<\lambda<1$ and $0<p<1$, 
$\tilde\lambda_{el}$ and $\tilde F_{2,el}(x)$ are not consistent estimators for $\lambda$ and $F_2(x)$. 

\subsection{General idea of binomial likelihood\label{binomial.malaria}}
Our method is motivated by the binomial likelihood approach proposed by \citet{Qin14}. 
We briefly review the main idea by considering a simple case. 
Suppose we have a random sample $Z_1,\ldots,Z_n$ from 
a population with cdf $G(x)$. 
The binomial likelihood is motivated by the fact that for any given $t$, 
$$
I(Z_i\leq t)\sim \text{Bin}\Big(1;G(t)\Big),
$$
where ``Bin" denotes the binomial distribution. 
Summing the log-likelihood of $\{I(Z_i\leq t)\}_{i=1}^n$ for the given $t$  
gives 
\begin{equation*}
\tilde l_B (G;t)=\sum_{i=1}^nI\left(Z_i\leq t\right)\log G(t)
+\left\{n- \sum_{i=1}^nI\left(Z_i\leq t\right) \right\}\log \bar G(t),
\end{equation*}
where $\bar G(t)=1-G(t)$.

We can arbitrarily choose the values of $t$ and then take the sum of $\tilde l_B (G;t)$ over the chosen values of $t$ or 
we can integrate $\tilde l_B(G;t)$ to get an objective function which does not depend on $t$. 
Let $z_1<\ldots<z_J$ be distinct values of the observed random sample $Z_1,\ldots,Z_n$. 
\citet{Qin14} suggested choosing the values of $t$ to be $\{z_j,~j=1,\ldots,J\}$ and then taking the sum of $\tilde l_B(G;t)$ over these $J$ values.  This gives the binomial likelihood as follows: 
$$
l_B(G)
=\sum_{j=1}^J\tilde l_B(G;z_j). 
$$
The maximum binomial likelihood estimator of $G(x)$ maximizes $l_B(G)$
subject 
to the constraint that $G(x)$ is a cdf. 
\citet{Qin14} show that  the empirical cdf of $X_1,\ldots,X_n$ is
a maximum binomial likelihood estimator of $G(x)$. 
They further apply this method to estimate the component cdfs of
mixture models with known proportions with censored data. 
\citet{Lee16} apply the  binomial likelihood method 
to estimate the distributional treatment effects in instrumental variable models.
To avoid dealing with potentially vanishing probabilities in the boundaries, 
they recommend using a subset of $z_1<\ldots<z_J$. 
They also establish the asymptotic properties of the maximum binomial likelihood estimator. 

Naturally, we may apply the binomial likelihood method to the data generated from Models \eqref{model1}
and \eqref{model2}. 
However, the binomial likelihood does not take the zero-inflated structure into account. 
Specifically, if we apply the binomial likelihood method 
to the data from Model \eqref{model1}, for any given $t>0$
$$
P(X_i\leq t)=p+(1-p) F_1(t),
$$
which implies that $p$ and $F_1$ are tangled together in the binomial likelihood. 
A similar problem occurs when the binomial likelihood method is applied to the data from  Model \eqref{model2}. 

\section{Maximum multinomial likelihood estimation for compound mixtures}
\label{method.malaria}

\subsection{Maximum multinomial likelihood}

Motivated by the idea of the binomial likelihood, 
we now propose a maximum multinomial likelihood to  
estimate $(\lambda,p,F_1,F_2)$ simultaneously. 
This method takes the zero-inflated and compound structure in (\ref{model1}) and (\ref{model2})
into account. 
We first introduce some notation. 
For any $t> 0$, 
let 
$$
{\bf M}_i(t)=\big(m_{i0}, m_{i1}(t), m_{i2}(t))^\tT=
\Big(
I(X_i=0), 
I(0<X_i\leq t), 
I(X_i>t) 
\Big)^\tT,~~i=1,\ldots,m
$$
and 
$$
{\bf N}_j(t)=\Big(n_{j0}, n_{j1}(t), n_{j2}(t))^\tT=
\big(
I(Y_j=0), 
I(0<Y_j\leq t), 
I(Y_j>t) 
\Big)^\tT,~~j=1,\ldots,n.
$$

The  multinomial likelihood method
is motivated by the fact that 
$$
{\bf M}_i(t)
\sim \text{Multi}\Big(1; p, (1-p) F_1(t), (1-p)\bar F_1(t)\Big),~~i=1,\ldots,m
$$
and 
$$
{\bf N}_j(t)
\sim \text{Multi}\Big(1; (1-\lambda)p, (1-\lambda)(1-p) F_1(t)+\lambda F_2(t), (1-\lambda)(1-p) \bar F_1(t)+\lambda \bar F_2(t)\Big), ~~j=1,\ldots,n,
$$
where ``Multi" denotes the multinomial distribution. 
Summing 
the log-likelihoods of $\left\{{\bf M}_i(t)\right\}_{i=1}^m$ and $\{ {\bf N}_j(t)\}_{j=1}^n$
gives 
\begin{eqnarray*}
\tilde l_M(\lambda, p, F_1,F_2; t)&=&\sum_{i=1}^m\left[m_{i0}\log p +m_{i1}(t)\log \{(1-p) F_1(t) \}+m_{i2}(t)\log \{(1-p)\bar F_1(t)\}\right]\\
&&+\sum_{j=1}^n\Big[ n_{j0}\log\{(1-\lambda)p\} +n_{j1}(t)\log \{(1-\lambda)(1-p) F_1(t)+\lambda F_2(t) \}\\
&&\hspace{0.5in}+ n_{j2}(t)\log \{(1-\lambda)(1-p) \bar F_1(t)+\lambda \bar F_2(t)\}\Big].
\end{eqnarray*}

Let $t_1<\ldots<t_k$ be the distinct values of the positive values in $\{X_1,\ldots,X_m\}$ and $\{Y_1,\ldots,Y_n\}$. 
For a given $q\in (0,0.5)$, define 
$$
I_q=\big\{\ceil{kq},\ceil{kq}+1,\ldots,\ceil{k(1-q)}\big\}, 
$$ 
 where $\ceil{x}$ is the smallest integer greater than or equal to $x$.
 For asymptotic purposes, 
we suggest choosing the values of $t$ as $\{t_h|h\in I_q\}$ and 
then taking the sum of $\tilde l_M(\lambda, p, F_1,F_2; t)$ over these values 
to obtain the  multinomial likelihood 
\begin{eqnarray}
\nonumber
l_M(\lambda, p, F_1,F_2)&=&\sum_{h\in I_q}\sum_{i=1}^m\left[m_{i0}\log p +m_{i1}(t_h)\log \{(1-p) F_1(t_h) \}+m_{i2}(t_h)\log \{(1-p)\bar F_1(t_h)\}\right]\\
\nonumber&&+\sum_{h\in I_q}\sum_{j=1}^n\Big[ n_{j0}\log\{(1-\lambda)p\} + n_{j1}(t_h)\log \{(1-\lambda)(1-p) F_1(t_h)+\lambda F_2(t_h) \}\\
&&\hspace{0.8in}+ n_{j2}(t_h)\log \{(1-\lambda)(1-p) \bar F_1(t_h)+\lambda \bar F_2(t_h)\}\Big].
\label{mmlh}
\end{eqnarray}
A more detailed discussion of 
the choice of the $t$ values will be given in Section \ref{malaria.section.theory}.

Further, let $k_q$ be the number of indexes in $I_q$ and 
$J(x,y)=x\log y+(1-x)\log(1-y)$, $x,y\in[0,1]$.
After some algebra work, the multinomial likelihood can be rewritten as  
\begin{align}
l_M(\lambda,p,F_1,F_2) =&mk_qJ\left(\tilde p,p\right)
+nk_q J\left((1-\tilde\lambda)\tilde p,(1-\lambda)p\right)\nonumber\\
& +\sum_{h\in I_q}\left\{m_+J\left(\tilde F_1(t_h) ,F_1(t_h)\right)
+n_+J\left(\tilde F_{Y+}(t_h),F_{Y+}(t_h)\right)\right\},
\label{ln.func0}
\end{align}
where $F_{Y+}(x)=(1-\lambda^*)F_1(x)+\lambda^*F_2(x)$, 
 $\tilde p$ and $\tilde\lambda$ are defined in (\ref{binest1.malaria}), and $\tilde F_1$ and $\tilde F_{Y+}$ are the empirical cdfs of the positive values in  $\{X_1,\ldots,X_m\}$
and $\{Y_1,\ldots,Y_n\}$, respectively.

The maximum multinomial likelihood estimator of $(\lambda, p, F_1,F_2)$ is defined as 
$$
(\hat\lambda, \hat p, \hat F_1,\hat F_2)=\arg\max_{(\lambda, p, F_1,F_2)\in \Theta}l_M(\lambda, p, F_1,F_2),
$$
where $\Theta=\{(\lambda, p, F_1,F_2)| \lambda, p\in[0,1], ~F_1\mbox{ and }F_2\mbox{ are cdfs}\}$.

\subsection{Asymptotic properties}
\label{malaria.section.theory}

The asymptotic properties of $(\hat\lambda, \hat p, \hat F_1,\hat F_2)$ rely on the following regularity conditions: 
\begin{enumerate}
\item[{\bf A1}.]  $0<\lambda<1$ and $0<p<1$.
\item[{\bf A2}.] $n/N=\rho\in(0,1)$, where $N=n+m$. 
\item[{\bf A3}.] 
$F_1(x)$ and $F_2(x)$ have the same support.
Further, $F_1(x)$ and $F_2(x)$ are absolutely continuous and strictly increasing on the support. 
\end{enumerate}

\begin{theorem}
\label{malaria.thm}
Suppose A1--A3 are satisfied. Then we have 
\begin{enumerate}
\item[(a)] $\hat{\lambda}-\lambda=O_p\left(N^{-1/2}\right),~~
\hat{p}-p=O_p\left(N^{-1/2}\right)$; 
\item[(b)] $k_q^{-1}\sum_{h\in I_q}\{ \hat{F}_1(t_h)-F_1(t_h)\}^2=O_p(N^{-1}),~~
k_q^{-1}\sum_{h\in I_q}\{ \hat{F}_2(t_h)-F_2(t_h)\}^2=O_p(N^{-1}).$
\end{enumerate}
\end{theorem}
 
 For convenience of presentation, the proof of Theorem \ref{malaria.thm} is in Appendix B.
 Here we make some comments on 
 the regularity conditions and the choice of the $t$ values in (\ref{mmlh}).
 
Condition A1 states that the true values of $\lambda$ and $p$ are interior points of their parameter space. 
Condition A2 requires that the sample size ratio in the two groups is a constant. 
Condition A3 and the choice of the $t$ values as $\{t_h: h\in I_q\}$
together ensure that $\tilde F_{Y+}(t_h)$ and $\tilde F_1(t_h)$ are uniformly away from 0 and 1 in probability for all $h\in I_q$. 
This guarantees that $l_M(\tilde \lambda,\tilde p,\tilde F_1,\tilde F_2^*)-
l_M(\hat \lambda,\hat p,\hat F_1,\hat F_2)$ can be bounded above and below by 
quadratic functions.
As a result, we can derive the asymptotic results in  Theorem \ref{malaria.thm}. 
A similar idea has been used by \citet{Lee16}, who 
use the maximum binomial method to estimate the distributional treatment effects in instrumental variable models.
 In practice, we recommend a small value for $q$; we used $q=0.001$ 
 in our simulation study and real-data analysis.

\subsection{EM-algorithm}
In (\ref{ln.func0}), $F_1(x)$ and $F_2(x)$ are tangled together,
which makes the explicit form of $(\hat\lambda, \hat p, \hat F_1,\hat F_2)$
unknown. 
Recall that $\lambda^*=\lambda/\left\{1-p(1-\lambda)\right\}$.  
Then 
$l_M(\lambda, p, F_1,F_2)$ can be rewritten as 
\begin{eqnarray}
l_M(\lambda, p, F_1,F_2)&=&k_q\left[m_0\log p+m_+\log(1-p)+n_0\log\{p(1-\lambda)\}+n_+\log\{1-p(1-\lambda)\}\right]\nonumber\\
&&+\sum_{h\in I_q}\sum_{i=1}^{m_+} \left\{m_{i1}(t_h)\log F_{1}(t_h)+  m_{i2}(t_h)\log \bar F_{1}(t_h)\right\}\nonumber\\
&&+\sum_{h\in I_q}\sum_{j=1}^{n_+}\log \Big[
(1-\lambda^*) \{ F_{1}(t_h)\}^{ n_{j1}(t_h)}\{\bar F_{1}(t_h)\}^{n_{j2}(t_h)}\nonumber\\
&&\hspace{1.1in} 
+\lambda^* \{ F_{2}(t_h)\}^{ n_{j1}(t_h)}\{\bar F_{2}(t_h)\}^{n_{j2}(t_h)} \Big],
\label{ln.func}
\end{eqnarray}
from which we have that 
\begin{equation}
\label{nj1.th}
n_{j1}(t_h)\sim (1-\lambda^*)\text{Bin}\Big(1,F_{1}(t_h)\Big)+ \lambda^*\text{Bin}\Big(1,F_{2}(t_h)\Big).
\end{equation}

With the mixture structure  above, 
the EM-algorithm naturally fits into our problem. We first define the missing data.
For $j=1,\ldots,n_+$ and $h\in I_q$, let 
$V_{jh}=1$ if  $n_{j1}(t_h)$ is from $\text{Bin}\Big(1,F_{2}(t_h)\Big)$, 
and 0 if $n_{j1}(t_h)$ is from $\text{Bin}\Big(1,F_{1}(t_h)\Big)$.
Further, let $ \mathcal{X} = \{X_1,\ldots,X_m,Y_1,\ldots,Y_n\}$ be the observed data,
$\mathcal{V}=\{V_{jh}|j=1,\ldots,n_+, h\in I_q\}$,
and $\bTheta = (\lambda, p, F_1,F_2)$.
Then, based on  $\{\mathcal{X},  \mathcal{V}\}$,
we can write down the complete multinomial likelihood 
and derive the EM-algorithm accordingly. 
For convenience of presentation, 
we leave the technical details to Appendix C and
present the E-step and M-step of the EM-iterations directly. 

Let $(\lambda^{(0)}, p^{(0)}, F_1^{(0)},F_2^{(0)})$ and $\lambda^{*(0)}$
 denote the initial values of $(\lambda, p, F_1,F_2)$ and $\lambda^*$, respectively.
 Further, 
we use $\bTheta^{(r)}=(\lambda^{(r)}, p^{(r)}, F_1^{(r)},F_2^{(r)})$
and $\lambda^{*(r)}$
to respectively denote the updated values of $\bTheta=(\lambda, p, F_1,F_2)$ and $\lambda^*$ after $r$ EM-iterations, $r=0,1,2,\ldots$.

In the E-step of the $r$th iteration, for $h=1,\ldots,k$ and $j=1,\ldots,n_+$, 
we calculate 
$$
E(V_{jh}|\mathcal{X},\bTheta^{(r-1)})
=\left\{ a_{h}^{(r)}\right\}^{ n_{j1}(t_h) }\left\{ b_{h}^{(r)}\right\}^{ n_{j2}(t_h)} ,
$$
where 
$$
a_{h}^{(r)}=\frac{ \lambda^{*(r-1)} F_2^{(r-1)}(t_h)} 
{
\lambda^{*(r-1)} F_2^{(r-1)}(t_h)+
(1-\lambda^{*(r-1)})F_1^{(r-1)}(t_h)
}
$$
and 
$$
b_{h}^{(r)}=
\frac{ \lambda^{*(r-1)}  \bar F^{(r-1)}_2(t_h)} 
{
\lambda^{*(r-1)} \bar F^{(r-1)}_2(t_h)+
(1-\lambda^{*(r-1)})\bar F_1^{(r-1)}(t_h)
}. 
$$ 

For any $t>0$, let 
$$
m_1(t)=\sum_{i=1}^m m_{i1}(t),~~ 
m_2(t)=\sum_{i=1}^m m_{i2}(t), ~~
 n_1(t)=\sum_{j=1}^n n_{j1}(t), \mbox{ and } n_2(t)=\sum_{j=1}^n n_{j2}(t). 
$$
In the M-step, 
we update $(\lambda, p, F_1,F_2)$ by 
\begin{eqnarray*}
\lambda^{(r)}&=&\frac{1}{k_qn}\sum_{h\in I_q}\left\{n_1(t_h)a_{h}^{(r)}+n_2(t_h)b_{h}^{(r)}\right\}
,\\
p^{(r)}&=&\frac{m_0+n_0}{m+n-n\lambda^{(r)}},\\
F_1^{(r)}&=&\arg\min_{F\mbox{ is a cdf}}\sum_{h\in I_q}
\left[
m_++n_1(t_h)\{1-a_h^{(r)}\}+n_2(t_h)\{1-b_h^{(r)}\}
 \right]
 \left[
\tilde F_1^{(r)}(t_h)
 -F(t_h)
 \right]^2,\\
F_2^{(r)}&=&\arg\min_{F\mbox{ is a cdf}}\sum_{h\in I_q}
\left\{
n_1(t_h) a_h^{(r)}+n_2(t_h)b_h^{(r)}
 \right\}
 \left\{
\tilde F_2^{(r)}(t_h)
 -F(t_h)
 \right\}^2,  
\end{eqnarray*}
where 
$$
\tilde F_1^{(r)}(t_h)
=
 \frac{m_1(t_h)+n_1(t_h)\{1-a_h^{(r)}\}}
 {m_++n_1(t_h)\{1-a_h^{(r)}\}+n_2(t_h)\{1-b_h^{(r)}\}}
\mbox{
 and 
 }
\tilde F_2^{(r)}(t_h)
=
 \frac{n_1(t_h)a_h^{(r)}}
 {n_1(t_h) a_h^{(r)}+n_2(t_h)b_h^{(r)}}.
 $$
 
Updating $F_1$ or $F_2$
is an isotonic regression problem, 
which can be  solved via the pool adjacent violators algorithm \citep{Ayer55}.
The E-step and M-step are iterated until convergence. 

We make two remarks about the above EM-algorithm.
Following the proof  in \citet{Dempster77}, 
we can show that the multinomial likelihood $l_M(\lambda, p, F_1,F_2)$ does not decrease after each iteration.
That is,  for $r\geq1$
$$
l_M ( \lambda^{(r)},p^{(r)},F_1^{(r)},F_2^{(r)})
\geq
l_M( \lambda^{(r-1)},p^{(r-1)},F_1^{(r-1)},F_2^{(r-1)}).
$$
Further, note that
$
l_M(\lambda, p, F_1,F_2)\leq 0
$
and $l_M(\lambda, p, F_1,F_2)$ is a continuous function of all the unknown parameters.
Then
the sequence
$\left\{ l_M ( \lambda^{(r)},p^{(r)},F_1^{(r)},F_2^{(r)}) \right\}_{r\geq1}$ eventually converges
to a stationary value of $l_M(\lambda, p, F_1,F_2)$  for a given initial value $\bTheta^{(0)}$ \citep{Wu83}.
However, this stationary value  may not be a global maximum. 
To increase the possibility  of finding the global maximum,
we recommend using multiple initial values. 
Our simulation  results demonstrate that this method works well.
In practice, we may stop the algorithm when the increment in $l_M ( \lambda^{(r)},p^{(r)},F_1^{(r)},F_2^{(r)})$
after an iteration is no greater than, say,
$10^{-6}$.

\section{Simulation study}
\label{simulation.malaria}

In this section, we perform a simulation study to compare the proposed maximum multinomial likelihood estimator $(\hat\lambda, \hat p, \hat F_1,\hat F_2)$ with $(\tilde\lambda,\tilde p,\tilde F_1,\tilde F_2)$.
Recall that the binomial estimator of $(\lambda,p)$ is defined in \eqref{binest1.malaria}
and 
the estimator $(\tilde F_1,\tilde F_2^*)$ of $(F_1,F_2)$ is defined in \eqref{binest2.malaria}. 
Since 
$\tilde{F}_2^*(x)$ may not be monotonically increasing, 
we apply isotonic regression to $\tilde{F}_2^*$ to obtain $\tilde F_2$: 
$$
\tilde F_2
=\arg\min_{F_2\mbox{ is a cdf}}~~\sum_{h=1}^{k} \{\tilde{F}_2^*(x)-F_2(x)\}^2,
$$
which is guaranteed to be an increasing function. 
We call $(\tilde{F}_1,\tilde{F}_2)$ a plug-in estimator of $(F_1,F_2)$. 
Note that the proposed estimator $(\hat\lambda, \hat p, \hat F_1,\hat F_2)$ and the estimator $(\tilde\lambda,\tilde p,\tilde F_1,\tilde F_2)$ are obtained under the same model assumptions.

We consider two scenarios as follows:  
\begin{itemize}
\item[]Scenario 1:  $F_1$ and $F_2$ are the cdfs of  $LN(0,1)$ and $LN(2,1)$ respectively,  where 
$LN(a, b)$ denotes a log-normal distribution with mean $a$ and variance $b$ both with respect to the log scale (i.e.~mean and variance of the associated normal random variable). 
\item[]Scenario 2:  $F_1$ and $F_2$ are the cdfs of  $GAM(1,1)$ and $GAM(4,3)$ respectively, 
where $GAM(a,b)$ denotes a gamma distribution
with shape parameter $a$ and scale parameter $b$.

\end{itemize}

For each scenario, we consider $(m,n)=(100,100)$ and $(m,n)=(150,250)$, 
and nine combinations of $(\lambda,p)$ with $\lambda,p\in\{0.25,0.5,0.75\}$.
For each combination of $(m,n,\lambda,p)$, the number of replications is 1000. 
The mean squared errors (MSEs) of the estimates of $\lambda$ and $p$
are used as the basis for comparing the different estimates of $\lambda$ and $p$. 
The Kolmogorov--Smirnov  distance between
the estimated and true cdfs
is used as a basis for comparing the different estimates of $F_1$ and $F_2$.
The simulation results are summarized in Tables \ref{table1.malaria}--\ref{table4.malaria}.

In each case, the proposed maximum multinomial likelihood estimate  $(\hat\lambda, \hat p, \hat F_1,\hat F_2)$ outperforms $(\tilde\lambda,\tilde p,\tilde F_1,\tilde F_2)$ 
as it gives more accurate estimates, especially when $\lambda$ is small.
The improvement of $(\hat\lambda,\hat F_2)$ over $(\tilde \lambda,\tilde F_2)$
is more obvious when $\lambda$ is small, when the information about the component with cdf $F_2$ is quite limited.

\section{Application to malaria data}
\label{example.malaria}
In this section, we analyze the data collected by \citet{Kitua96}. 
Similarly to \citet{Qin05}, we consider a subset of this dataset for children aged between six and nine months collected in January--June, the wet season,  and in July--December, the dry season. 
The data sets are downloadable from {\tt http://www.blackwellpublishers.co.uk/rss}.
The measurements are the parasite levels per $\mu l$, ranging from 0 to $399952.1$.
Among these measurements, there are $n=264$ observations
from the mixture sample (i.e.~from the wet season) and
$m=144$ observations  for nonmalaria individuals
from the dry season.

We apply the proposed estimator $(\hat\lambda, \hat p, \hat F_1,\hat F_2)$ 
 along with $(\tilde\lambda,\tilde p,\tilde F_1,\tilde F_2)$ to the data above.
 To evaluate the variability of these two estimators, 
 we use the nonparametric bootstrap method \citep{Efron81} with 1000 bootstrap samples
 to obtain bootstrap standard errors (SEs) of the estimators, 
 bootstrap percentile confidence intervals (BPCIs) for $\lambda$ and $p$, and 
bootstrap confidence bands for $F_1$ and $F_2$. 
The results for $\lambda$ and $p$ are summarized in 
Table \ref{example.table}. 
Clearly, the bootstrap SEs of $\hat\lambda$ and $\hat p$
are smaller than those of $\tilde\lambda$ and $\tilde p$. 
The 95\% BPCIs for $\lambda$ and $p$
based on the proposed method 
have shorter lengths than those based on the binomial estimator. 
This indicates that our method is more efficient than the binomial estimator, 
which is in accordance with the simulation study.

 The 95\% bootstrap pointwise confidence bands of $ {F}_1$ and $F_2$ are shown in Figure \ref{cdf.malaria}.
 We see that 
the confidence bands for $F_1$ based on our method and the plug-in method are quite close to each other. 
However, the confidence bands for $F_2$ based on our method are narrower than those 
based on the plug-in method, especially when the parasite level is smaller than 150000. 
This indicates that our method is more efficient than the plug-in method in terms of estimating $F_2$.

For an individual in the mixture sample, set $D=1$ if the corresponding subject is infected with malaria and set $D= 0$ otherwise.
Let $g_1(x)$ and $g_2(x)$ be the pdfs of the logarithm of the positive parasite levels
in the nonmalaria and malaria populations, respectively. 
Then given the log parasite level $x$, 
the posterior probability of $D=1$
is given by
$$
\eta(x)= \frac{\lambda^*g_2(x)}{(1-\lambda^*)g_1(x)+\lambda^*g_2(x)}. 
$$
We can use $\hat F_1$ and $\hat F_2$ to construct an estimator for $\eta(x)$. 

We first estimate $g_1(x)$ and $g_2(x)$
as 
$$
\hat g_1(x)=\int K_{h_1}(\log t-x) d\hat F_1(t), ~~\hat g_2(x)=\int K_{h_2}(\log t-x) d\hat F_2(t),
$$
where $K_h(x)=K(x/h)/h$, $K(x)$ is a symmetric kernel, and $h$ is the bandwidth. 
For illustration, we use the standard normal density function for $K(x)$ 
and 
 choose the bandwidth $h$ by rule of thumb:
$h_1= 1.06 \hat \sigma_1  (m_++n_+)^{-1/5}$ and $h_2= 1.06 \hat \sigma_2  n_+^{-1/5}$, 
where 
$$
\hat\sigma_1^2=\int_{x}(\log x)^2d\hat F_1(x)- \left\{\int_{x}\log x d\hat F_1(x) \right\}^2,~~
\hat\sigma_2^2=\int_{x}(\log x)^2d\hat F_2(x)- \left\{\int_{x} \log x d\hat F_2(x) \right\}^2.
$$
Since all the positive observations in $\{X_1,\ldots,X_m,Y_1,\ldots,Y_n\}$ are used to estimate $F_1$ and only the positive observations in $\{Y_1,\ldots,Y_n\}$ are used to estimate $F_2$, 
$m_++n_+$ is used in $h_1$ and $n_+$ is used in $h_2$.

Figure \ref{den.malaria} shows the density estimate of the nonmalaria population (i.e.~observations from the dry season), the density estimate of the mixture sample (i.e.~observations from the wet season), 
and the estimated posterior probability of catching malaria given the logarithm of the positive parasite level in the mixture sample. 
We can see that the density estimates match the histograms well for the data from both the dry season and the wet season. 
This indicates that the density estimators $\hat g_1$ and $\hat g_2$ work well for the malaria data.
Further, we can see that the estimated posterior probability of catching malaria given the logarithm of the positive parasite level is an increasing function of the log parasite level. 
This agrees with clinical malaria knowledge \citep{Vounatsou98}: the higher the parasite level, the higher the probability that an individual is infected with malaria. 

\section{Conclusion}
\label{conclusion.malaria}
We have proposed a maximum multinomial likelihood method 
to estimate the unknown parameters in a compound mixture model \eqref{model2}. 
We established the asymptotic properties of the proposed estimator and 
developed an EM-algorithm to numerically calculate the estimate. 
Our method is more efficient than the simple and natural estimator $(\tilde\lambda,\tilde p,\tilde F_1,\tilde F_2)$. 
We illustrate our method using a malaria dataset. 

We are not currently able to establish the limiting distribution of our estimator. 
Further, the asymptotic properties of $\hat f_1$ and $\hat f_2$ will help us to understand the properties 
of the estimated posterior probability $\eta(x)$. 
We leave these topics to future work.

\section*{Disclosure statement}

 No potential conflict of interest was reported by the authors.

\section*{Funding}

Drs. Liang and Li's work are partially supported by the
Natural Sciences and Engineering Research Council of Canada. 
%

%
\bibliographystyle{agsm}

\bibliography{reff}

\section*{Appendix A: Proof of Proposition 1}
For a given vector $\ba$ such that $a_i\geq0$ and $\sum_{i=1}^{m_++n_+}a_i=1$, 
let $a_+=\sum_{i=1}^{m_+} a_i$, 
$$
a_{i}^*=\frac{a_+}{m_+},~~i=1,\ldots,m_+, ~~a_{j+m_+}^*=\frac{1-a_+}{n_+}, ~~j=1,\ldots, n_+,
$$
and 
$
\ba^*=(a_1^*,\ldots,a_{m_++n_+}^*)^\tT. 
$
Further, let $\bb^*=(1/n_+,\ldots,1/n_+)^\tT$. 

Using Jensen's inequality, we can show that 
\begin{equation}
\label{el.ine.malaria}
\tilde l_{el}(\lambda,p,\mathbf{a},\mathbf{b})
\leq 
\tilde l_{el}(\lambda,p,\mathbf{a}^*,\mathbf{b}^*)
\end{equation}
with equality holding if and only if $\ba=\ba^*$ and $\bb=\bb^*$. 
Hence, $\tilde b_{j,el}=\bb^*$. 
The inequality in (\ref{el.ine.malaria}) also implies that 
we can concentrate on the case where $\bb=\bb^*$ and
$$
a_1=\ldots=a_{m_+}=a,~~a_{m_++1}=\ldots=a_{m_++n_+}=a^*,
$$
where $a\geq 0$, $a^*\geq 0$, and $m_+a+n_+a^*=1$. 
In this case, $\tilde l_{el}(\lambda,p,\mathbf{a},\mathbf{b})$ becomes 
\begin{align*}
\tilde l_{el}^*(\lambda,p,a,a^*)&=m_0\log p+m_+\log(1-p)+m_+\log a\\
&+n_0\log(1-\lambda)+n_0\log p+n_+\log\left\{(1-\lambda)(1-p)a^*+\lambda/n_+\right\}
\end{align*}
and 
the parameter space for $(\lambda,p,a,a^*)$
 is
$$
\Theta^*=\{(\lambda,p,a,a^*)| \lambda\in[0,1],~p\in[0,1],~a\in[0,1/m_{+}],~a^*\in[0,1/n_+],~~m_+a+n_+a^*=1\}.
$$ 
Let 
$$
(\tilde \lambda_{el},\tilde p_{el},\tilde a_{el},\tilde a_{el}^*)=\arg\max_{(\lambda,p,a,a^*)\in\Theta} \tilde l_{el}^*(\lambda,p,a,a^*). 
$$

Define 
\begin{align*}
\Theta^*_1&=\{(\lambda,p,a,a^*)\in\Theta^*| \lambda\in(0,1),~p\in(0,1),~a\in (0,1/m_{+})\},\\
\Theta^*_2&=\{(\lambda,p,a,a^*)\in\Theta^*| a=1/m_+\},\\
\Theta^*_3&=\{(\lambda,p,a,a^*)\in\Theta^*| \lambda=0\}.
\end{align*}

Clearly, $\tilde \lambda_{el}\neq 1$ and $\tilde a_{el}\neq 0$, $\tilde p_{el}\neq$ 0 or 1. 
Thus, $(\tilde \lambda_{el},\tilde p_{el},\tilde a_{el},\tilde a_{el}^*)$ belongs
to $\Theta^*_1$, $\Theta^*_2$, or $\Theta^*_3$. 
To complete the proof, we need to argue that $(\tilde \lambda_{el},\tilde p_{el},\tilde a_{el},\tilde a_{el}^*)\in \Theta^*_2$.  We first argue that $(\tilde \lambda_{el},\tilde p_{el},\tilde a_{el},\tilde a_{el}^*)\notin \Theta^*_1$.
 
Define
\begin{align*}
\tilde{L}_{el}^*(\lambda,p,a,a^*)=&m_0\log p+m_+\log(1-p)+m_+\log a\nonumber+n_0\log(1-\lambda)+n_0\log p\\
&+n_+\log\left\{(1-\lambda)(1-p)a^*+\lambda/n_+\right\}+s(m_+a+n_+a^*-1),
\end{align*}
where $s$ is the Lagrange multiplier. 
If $(\tilde \lambda_{el},\tilde p_{el},\tilde a_{el},\tilde a_{el}^*)\in \Theta^*_1$, 
then $(\tilde \lambda_{el},\tilde p_{el},\tilde a_{el},\tilde a_{el}^*)$ will satisfy 
\begin{align}
\label{ellem4}
\frac{\partial \tilde{L}_{el}^*(\tilde \lambda_{el},\tilde p_{el},\tilde a_{el},\tilde a_{el}^*) } {\partial \lambda}&=-\frac{n_0}{1-\tilde\lambda_{el}}+\frac{1-n_+(1-\tilde{p}_{el})\tilde{a}_{el}^*}{(1-\tilde\lambda_{el})(1-\tilde{p}_{el})\tilde{a}_{el}^*+\tilde\lambda_{el}/n_+}=0,\\
\label{ellem5}
\frac{\partial \tilde{L}_{el}^*(\tilde \lambda_{el},\tilde p_{el},\tilde a_{el},\tilde a_{el}^*) } {\partial p}&=\frac{m_0+n_0}{\tilde{p}_{el}}-\frac{m_+}{1-\tilde{p}_{el}}-\frac{n_+(1-\tilde\lambda_{el})\tilde{a}_{el}^*}{(1-\tilde\lambda_{el})(1-\tilde{p}_{el})\tilde{a}_{el}^*+\tilde\lambda_{el}/n_+}=0,\\
\label{ellem6}
\frac{\partial \tilde{L}_{el}^*(\tilde \lambda_{el},\tilde p_{el},\tilde a_{el},\tilde a_{el}^*) } {\partial a}&=\frac{m_+}{\tilde{a}_{el}}+sm_+=0,\\
\label{ellem7}
\frac{\partial \tilde{L}_{el}^*(\tilde \lambda_{el},\tilde p_{el},\tilde a_{el},\tilde a_{el}^*) } {\partial a^*}&=\frac{n_+(1-\tilde\lambda_{el})(1-\tilde{p}_{el})}{(1-\tilde\lambda_{el})(1-\tilde{p}_{el})\tilde{a}_{el}^*+\tilde\lambda_{el}/n_+}+sn_+=0.
\end{align}
Further, 
\begin{equation}
\label{ellem8}
m_+\tilde a_{el}+n_+\tilde a^*_{el}-1=0.
\end{equation}
Combining (\ref{ellem6}), (\ref{ellem7}), and (\ref{ellem8}) leads to
\begin{align}
\label{ellem9}
\tilde{a}_{el}^*&=\frac{n_+(1-\tilde\lambda_{el})(1-\tilde{p}_{el})-m_+\tilde\lambda_{el}}{n_+(m_++n_+)(1-\tilde\lambda_{el})(1-\tilde{p}_{el})}.
\end{align}
By (\ref{ellem4}), we have 
\begin{equation}
\label{ellem10}
(1-\tilde\lambda_{el})(1-\tilde{p}_{el})\tilde{a}_{el}^*+\tilde\lambda_{el}/n_+=1/n,
\end{equation}
which, together with (\ref{ellem9}), results in
\begin{equation}
\label{ellem11}
(1-\tilde\lambda_{el})(1-\tilde{p}_{el})+\tilde\lambda_{el}=\frac{m_++n_+}{n}.
\end{equation}
Substituting  (\ref{ellem10}) into  (\ref{ellem5}), 
we get
\begin{equation}
\label{ellem12}
(m+n)\tilde{p}_{el}=m_0+n_0+n \tilde\lambda_{el}.
\end{equation}
Combining (\ref{ellem11}) and (\ref{ellem12}), we have 
$
\tilde{p}_{el}=1,
$
which is a contradiction. 
Hence, $(\tilde \lambda_{el},\tilde p_{el},\tilde a_{el},\tilde a_{el}^*)\notin \Theta^*_1$ 
and 
$$
(\tilde \lambda_{el},\tilde p_{el},\tilde a_{el},\tilde a_{el}^*)\in \Theta^*_2\cup \Theta^*_3.
$$

We can show that 
\begin{align*}
\max_{(\lambda,p,a,a^*)\in \Theta^*_2}\tilde l^*_{el}(\lambda,p,a,a^*)=&(m_0+n_0)\log(m_0+n_0)+n_0\log n_0-n\log n-(m+n_0)\log(m+n_0),\\
\max_{(\lambda,p,a,a^*)\in \Theta^*_3}\tilde l^*_{el}(\lambda,p,a,a^*)=&(m_0+n_0)\log(m_0+n_0)-(m+n)\log(m+n).
\end{align*}
It can be checked that 
\begin{equation*}
n\log n+(m+n_0)\log(m+n_0)< n_0\log n_0+(m+n)\log(m+n),
\end{equation*}
which implies that  
$$
\max_{(\lambda,p,a,a^*)\in \Theta^*_2}\tilde l^*_{el}(\lambda,p,a,a^*)
>\max_{(\lambda,p,a,a^*)\in \Theta^*_3}\tilde l^*_{el}(\lambda,p,a,a^*).
$$
Therefore, $(\tilde \lambda_{el},\tilde p_{el},\tilde a_{el},\tilde a_{el}^*)\in \Theta^*_2$. 

Note that 
$$
\arg\max_{(\lambda,p,a,a^*)\in \Theta^*_2} \tilde l^*_{el}(\lambda,p,a,a^*)
=
\left(\frac{n_+}{n},\frac{m_0+n_0}{m_0+n_0+m_+},\frac{1}{m_+},0\right).
$$
Then 
\begin{align*}
\tilde{\lambda}_{el}&=\frac{n_+}{n},\\
\tilde{p}_{el}&=\frac{m_0+n_0}{m_0+n_0+m_+},\\
\tilde{a}_{el,i}&=\tilde a_{el}=\frac{1}{m_+},~i=1,\cdots,m_+,\\
\tilde{a}_{el,i}&=\tilde a_{el}^*=0,~i=m_++1,\cdots,m_++n_+,\\
\tilde{b}_{el,j}&=\frac{1}{n_+},~j=1,\cdots,n_+.
\end{align*}
This completes the proof of Proposition 1. 

\section*{Appendix B: Proof of Theorem 1}
\subsection*{Technical preparation}
Since both $F_1$ and $F_2$ are absolutely continuous, 
without loss of generality, 
we assume throughout the proofs that there are no ties in the positive values of  $\{X_i\}_{i=1}^m$ and $\{Y_j\}_{j=1}^n$.
Further, define 
$$
H_{k}(x)=\frac{1}{k}
\sum_{h=1}^kI(t_h\leq x).
$$
Denote 
$
t_L=\min\{t_h:~h\in I_q\}
$ 
and 
$
t_U=\max\{t_h:~h\in I_q\}.
$
The following lemma summarizes some useful results for the proof of Theorem 1.

\begin{lemma}
Let $
H(x)=(1-\eta) F_1(x)+\eta F_2(x),
$
where 
$$
\eta=\frac{\rho \lambda}{(1-\rho)(1-p)+\rho(1-\lambda)(1-p)+\rho\lambda}. 
$$
Under Conditions A1--A3, we have
\begin{enumerate}
\item[(a)] $\tilde\lambda-\lambda=O_p(N^{-1/2})$, $\tilde p-p=O_p(N^{-1/2})$, $\tilde \lambda ^*-\lambda^*=O_p(N^{-1/2})$; 
\item[(b)] 
$t_L\to H^{-1}(q)$ and $t_U\to H^{-1}(1-q)$ in probability;
\item[(c)] 
$\sup_{x}|\tilde F_1(x)-F_1(x)|=O_p(N^{-1/2})$ and
$\tilde F_1(t_L)\to F_1\big(H^{-1}(q)\big)$ and $\tilde F_1(t_U)\to F_1\big(H^{-1}(1-q)\big)$ in probability; 
\item[(d)] 
$\sup_{x}|\tilde F_2^*(x)-F_2(x)|=O_p(N^{-1/2})$ and
$\tilde F_2^*(t_L)\to F_2\big(H^{-1}(q)\big)$ and $\tilde F_2^*(t_U)\to F_2\big(H^{-1}(1-q)\big)$ in probability.

\end{enumerate}
\label{malaria.lem1}
\end{lemma}
\proof
We start with Part (a). 
Recall that 
$$
\tilde{p}=\frac{m_0}{m},~\tilde{\lambda}=1-\frac{n_0}{n\tilde{p}}.
$$
By classical asymptotic theory \citep{serfling2000} and Conditions A1--A2, 
we have 
$$
\tilde p-p=O_p(N^{-1/2})\mbox{ and }
(1-\tilde\lambda) \tilde p-(1-\lambda)p=O_p(N^{-1/2}),
$$
which together with the delta method imply that $\tilde \lambda-\lambda=O_p(N^{-1/2})$ and $\tilde \lambda ^*-\lambda^*=O_p(N^{-1/2})$. 
This completes Part (a).

Next we consider Part (b). 
Note that 
$$
t_L=H^{-1}_k(q)~~\mbox{ and }~~
t_U=H^{-1}_k(1-q). 
$$
To establish the  consistency of $t_L$ and $t_U$, 
we first argue that 
\begin{equation}
\label{sup.H}
\sup_{x}|H_k(x)-H(x)|=O_p(N^{-1/2}). 
\end{equation} 

Let 
$$
F_{X,m}(x)=\frac{1}{m}\sum_{i=1}^m I(X_i\leq x),~~
F_{Y,n}(x)=\frac{1}{n}\sum_{j=1}^n I(Y_j\leq x).
$$
Then for $x>0$, 
\begin{align*}
H_k(x)&=\frac{1}{k}\left[
m \{F_{X,m}(x)-F_{X,m}(0)\}
+
n\{F_{Y,n}(x)-F_{Y,n}(0)\}
\right]\\
&=\frac{1}{k/(n+m)} \left[
(1-\rho) \{F_{X,m}(x)-F_{X,m}(0)\}
+
\rho\{F_{Y,n}(x)-F_{Y,n}(0)\}\right].
\end{align*}
Note that $H(x)$ can be rewritten as 
$$
H(x)
=\frac{1}{(1-\rho)(1-p)+\rho\{(1-\lambda)(1-p)+\lambda\}}
\left[(1-\rho)\{F_{X}(x)-F_{X}(0) \}+\rho \{F_{Y}(x)-F_{Y}(0)\}\right].
$$
By the classical central limit theorem, the delta method, and Condition A2, 
\begin{equation}
\label{lemma1.part11}
k/(n+m)-[(1-\rho)(1-p)+\rho\{(1-\lambda)(1-p)+\lambda\}] =O_p(N^{-1/2}).
\end{equation}
By the triangular inequality, 
\begin{eqnarray}
\sup_x|H_k(x)-H(x)|&\leq&\left|\frac{1}{k/(n+m)}-\frac{1}{(1-\rho)(1-p)+\rho\{(1-\lambda)(1-p)+\lambda\}} \right| \nonumber\\
&&+\frac{2\sup_x|F_{X,m}(x)-F_{X}(x)|+2\sup_x|F_{Y,n}(x)-F_{Y}(x)|}{(1-\rho)(1-p)+\rho\{(1-\lambda)(1-p)+\lambda\}}.\label{lemma1.part13} 
\end{eqnarray}
By Conditions A1--A2, the uniform convergence rate of empirical cdfs, and the delta method, 
(\ref{lemma1.part11}) and (\ref{lemma1.part13}) lead to 
$$
\sup_x|H_k(x)-H(x)|=O_p(N^{-1/2}). 
$$
This, together with Condition A3, implies that $t_L\to H^{-1}(q)$ and $t_U\to H^{-1}(1-q)$ in probability. This completes Part (b). 

We now consider (c). 
Similarly to (\ref{sup.H}), we have 
$$
\sup_x|\tilde F_1(x)-F_1(x)|=O_p(N^{-1/2}),
$$
which implies that 
$
\tilde F_1(t_L)-F_1(t_L)\to0
$
in probability. 
Further, by the continuous mapping theorem, Part (a),  and Condition A3, we have 
$$
F_1(t_L)-F_1\big(H^{-1}(q)\big)
\to 0
$$
in probability. Hence, 
$$
\tilde F_1(t_L)-F_1\big(H^{-1}(q)\big) \to 0
$$
in probability. 
Similarly, we can establish the consistency of $\tilde F_1(t_U)$. This completes Part (c). 

Finally, the proof of Part (d) is similar to that of Part (c), so it is omitted. 
\qed

\subsection*{Proof of Theorem 1}
For convenience of presentation, we define more notation. 
Let 
$$
\hat F_{Y+}=(1-\hat\lambda^*) \hat F_1(x)+\hat\lambda^*\hat F_2(x), 
$$
where 
$$
\hat \lambda^*=\frac{\hat\lambda}{1-\hat p+\hat p\hat \lambda}. 
$$
For $x\in[t_L,t_U]$,
let 
$$
\breve F_2^*(x)=\arg\min_{F_2(x)\mbox{ is increasing}}~~\sum_{h\in I_q} \{\tilde{F}_2^*(x)-F_2(x)\}^2
$$
and 
$$
\breve F_2(x)=\min[ \max \{\breve F_2^*(x),0 \},1].
$$
Further, define
$$
\breve F_{Y+}(x)=(1-\tilde \lambda^*) \tilde F_1(x)+\tilde \lambda^*\breve F_2(x). 
$$

Let 
$$
A_{n,m}=\{\tilde p\in(0,1), \tilde \lambda^*\in(0,1), ~
0<\tilde F_2^*(t_h)<1,~
h\in I_q\}. 
$$
By Lemma \ref{malaria.lem1} and Condition A1, $P(0<\tilde p<1)\to 1$, $P(0<\tilde\lambda<1)\to 1$, and $P(0<\tilde F_2^*(t_h)<1,~~h\in I_q)\to 1$ as $N\to\infty$.
We therefore concentrate on the sample points in $A_{n,m}$.
Since $\breve F_2^*(x)$ is the isotonic regression function of $\tilde F_2^*(x)$,
we have $\max_{h\in I_q}\breve F_2^*(t_h)\leq\max_{h\in I_q}\tilde F_2^*(t_h)$ 
and $\min_{h\in I_q}\breve F_2^*(t_h)\geq\min_{h\in I_q}\tilde F_2^*(t_h)$.
As a result, for every sample point in $A_{n,m}$, 
we also have 
$\breve F_2^*(t_h)=\breve F_2(t_h)$. 

The roadmap for Theorem 1 is as follows. 
In the first step, we find a lower bound for
$
l_M(\tilde \lambda,\tilde p,\tilde F_1,\tilde F_2^*)-
l_M(\hat \lambda,\hat p,\hat F_1,\hat F_2), 
$
which is a quadratic function of $\hat{p}-\tilde{p}$, $(1-\hat{\lambda})\hat{p}-(1-\tilde{\lambda})\tilde{p}$, $\hat F_1-\tilde F_1$, and $\hat F_{Y+}-\tilde F_{Y+}$. 
In the second step, we argue that $l_M(\tilde \lambda,\tilde p,\tilde F_1,\tilde F_2^*)-
l_M(\hat \lambda,\hat p,\hat F_1,\hat F_2)$ is bounded above by a $O_p(N)$ term. 
The results then follow. 

For the first step, note that if $x$ is a fixed number, 
then \citep{Lee16}
\begin{equation}
\label{malaria.ine}
J(x,x)-J(x,y)\geq 0.5(x-y)^2. 
\end{equation}
Hence,
\begin{eqnarray}
&&l_M(\tilde \lambda,\tilde p,\tilde F_1,\tilde F_2^*)-
l_M(\hat \lambda,\hat p,\hat F_1,\hat F_2)\nonumber\\
&\geq&
\frac{1}{2}mk_q(\hat{p}-\tilde{p})^2
+\frac{1}{2}nk_q\left\{(1-\hat{\lambda})\hat{p}-(1-\tilde{\lambda})\tilde{p}\right\}^2\nonumber
\\
&&+\frac{m_+}{2}\sum_{h\in I_q}\left\{\hat{F}_1(t_h)-\tilde{F}_1(t_h)\right\}^2+\frac{n_+}{2}\sum_{h\in I_q}\left\{\hat{F}_{Y+}(t_h)-\tilde{F}_{Y+}(t_h)\right\}^2.
\label{malaria.thm.lowerbound1}
\end{eqnarray}

For the second step, we find an upper bound for $l_M(\tilde \lambda,\tilde p,\tilde F_1,\tilde F_2^*)-
l_M(\hat \lambda,\hat p,\hat F_1,\hat F_2)$. 
Since $(\hat \lambda,\hat p,\hat F_1,\hat F_2)$ is the maximum multinomial likelihood estimator, we have
$$
l_M(\tilde \lambda,\tilde p,\tilde F_1,\tilde F_2^*)-
l_M(\hat \lambda,\hat p,\hat F_1,\hat F_2)
\leq l_M(\tilde \lambda,\tilde p,\tilde F_1,\tilde F_2^*)-
l_M(\tilde \lambda,\tilde p,\tilde F_1,\breve F_2).
$$
Therefore, it suffices to find an upper bound for $l_M(\tilde \lambda,\tilde p,\tilde F_1,\tilde F_2^*)-
l_M(\tilde \lambda,\tilde p,\tilde F_1,\breve F_2)$. 

Let $a\wedge b=\min(a,b)$ and $a\vee b=\max(a,b)$. 
Applying the first-order Taylor expansion, we have 
\begin{align*}
&l_M(\tilde \lambda,\tilde p,\tilde F_1,\tilde F_2^*)-
l_M(\tilde \lambda,\tilde p,\tilde F_1,\breve F_2)
\notag\\
=&\sum_{h\in I_q}n_+\left\{J\big(\tilde{F}_{Y+}(t_h),\tilde{F}_{Y+}(t_h)\big)-J\big(\tilde{F}_{Y+}(t_h),\breve{F}_{Y+}(t_h)\big)\right\}
\notag\\
=&\sum_{h\in I_q}n_+\left\{\frac{\tilde{F}_{Y+}(t_h)}{\delta^2(t_h)}+\frac{1-\tilde{F}_{Y+}(t_h)}{\{1-\delta(t_h)\}^2}\right\}
                 \left\{\tilde{F}_{Y+}(t_h)-\breve{F}_{Y+}(t_h)\right\}^2, 
\end{align*}
where $\delta(t_h)\in[\tilde{F}_{Y+}(t_h)\wedge\breve{F}_{Y+}(t_h),\tilde{F}_{Y+}(t_h)\vee\breve{F}_{Y+}(t_h)]$. 

By the definition of $\tilde{F}_{Y+}(x)$ and $\breve{F}_{Y+}(x)$, we further have 
for every sample point in $A_{n,m}$, 
$$
\tilde F_{Y+}(t_h)\geq(1-\tilde\lambda^*)\tilde F_1(t_L),~~
\breve F_{Y+}(t_h)\geq(1-\tilde\lambda^*)\tilde F_1(t_L),
$$
and 
$$
1-\tilde F_{Y+}(t_h)\geq(1-\tilde\lambda^*)\left\{1-\tilde F_1(t_U)\right\},~~
1-\breve F_{Y+}(t_h)\geq(1-\tilde\lambda^*)\left\{1-\tilde F_1(t_U)\right\}.
$$
Therefore, 
\begin{align}
l_M(\tilde \lambda,\tilde p,\tilde F_1,\tilde F_2^*)-
l_M(\tilde \lambda,\tilde p,\tilde F_1,\breve F_2)
\leq&\frac{n_+\left(\tilde\lambda^*\right)^2}{(1-\tilde\lambda^*)^2 [\tilde F_1(t_L)\{1-\tilde F_1(t_U)\}]^2}  \sum_{h\in I_q}
                 \left\{\tilde{F}_{2}^*(t_h)-\breve{F}_{2}(t_h)\right\}^2                 \nonumber \\
                 =&O_p(N) \sum_{h\in I_q}
                 \left\{\tilde{F}_{2}^*(t_h)-\breve{F}_{2}^*(t_h)\right\}^2\nonumber\\
                 \leq& O_p(N) \sum_{h\in I_q}
                 \left\{\tilde{F}_{2}^*(t_h)-F_{2}(t_h)\right\}^2\nonumber\\
                 =&O_p(N),\label{malaria.thm.upperbound1}
\end{align}
where the last three steps follow from Lemma \ref{malaria.lem1} and $\breve{F}_{2}^*(t_h)=\breve{F}_{2}(t_h)$ for any sample point in $A_{n,m}$. 

Combining (\ref{malaria.thm.lowerbound1}) and (\ref{malaria.thm.upperbound1}), 
we have 
\begin{eqnarray}
mk_q(\hat{p}-\tilde{p})^2&=&O_p(N)\label{malaria.thm.last1},\\
nk_q\left\{(1-\hat{\lambda})\hat{p}-(1-\tilde{\lambda})\tilde{p}\right\}^2&=&O_p(N), \label{malaria.thm.last2}\\
\frac{m_+}{2}\sum_{h\in I_q}\left\{\hat{F}_1(t_h)-\tilde{F}_1(t_h)\right\}^2&=&O_p(N),\label{malaria.thm.last3}\\
\frac{n_+}{2}\sum_{h\in I_q}\left\{\hat{F}_{Y+}(t_h)-\tilde{F}_{Y+}(t_h)\right\}^2
&=&O_p(N). \label{malaria.thm.last4}
\end{eqnarray}
By Conditions A1 and A2, $k_q$, $m$, $n$, $m_+$, and $n_+$ all have the same order as $N$.
Hence, (\ref{malaria.thm.last1}) and (\ref{malaria.thm.last2}) lead to
$$
\hat{p}-\tilde{p}=O_p(N^{-1/2}),~~(1-\hat{\lambda})\hat{p}-(1-\tilde{\lambda})\tilde{p}=O_p(N^{-1/2}),
$$
which together with Part (a) of Lemma \ref{malaria.lem1}  implies 
$$
\hat{p}-p=O_p(N^{-1/2}),~~
\hat{\lambda}-\lambda=O_p(N^{-1/2}).
$$
Similarly, (\ref{malaria.thm.last3})  implies 
$$
\sum_{h\in I_q}\left\{\hat{F}_1(t_h)-\tilde{F}_1(t_h)\right\}^2=O_p(1). 
$$
By Part (c) of Lemma \ref{malaria.lem1} and the triangular inequality, 
we easily conclude that
\begin{equation}
\label{malaria.hatF1}
\sum_{h\in I_q}\left\{\hat{F}_1(t_h)-{F}_1(t_h)\right\}^2=O_p(1).
\end{equation}
Since $k_q$ has the same order as $N$, we further have 
  $$
\frac{1}{k_q}\sum_{h\in I_q}\left\{\hat{F}_1(t_h)-{F}_1(t_h)\right\}^2=O_p(N^{-1}).
$$
Similarly to (\ref{sup.H}), we have 
$$
\sup_x|\tilde F_{Y+}(x)-F_{Y+}(x)|=O_p(N^{-1/2}),
$$
which together with (\ref{malaria.thm.last4}) implies that 
$$
\sum_{h\in I_q}\left\{\hat{F}_{Y+}(t_h)-{F}_{Y+}(t_h)\right\}^2=O_p(1). 
$$
With $
\hat{p}-p=O_P(N^{-1/2}),~~
\hat{\lambda}-\lambda=O_P(N^{-1/2}),
$ and the delta method, 
we have $\hat\lambda^*-\lambda^*=O_p(N^{-1/2})$. 
By the triangular inequality and the form of $\hat{F}_{Y+}(t_h)$, we get
$$
\sum_{h\in I_q}\left\{ (1-\hat\lambda^*) \hat F_1(t_h)+\hat\lambda^* \hat F_2(t_h) -(1-\hat\lambda^*) F_1(t_h)-\hat\lambda^* F_2(t_h)\right\}^2=O_p(1). 
$$
By (\ref{malaria.hatF1}) and the triangular inequality, we obtain
$$
(\hat\lambda^*)^2
\sum_{h\in I_q}\left\{ \hat F_2(t_h) -F_2(t_h)\right\}^2=O_p(1), 
$$
which, together with $$\hat\lambda^*-\lambda^*=O_p(N^{-1/2})$$ and the fact that $k_q$ has the same order as $N$, 
implies 
\begin{equation*}
\label{malaria.hatF2}
\sum_{h\in I_q}\left\{\hat{F}_2(t_h)-{F}_2(t_h)\right\}^2=O_p(1).
\end{equation*}
This completes the proof of Theorem 1.

\section*{Appendix C: Details of EM-algorithm}
Based on $\{\mathcal{X},  \mathcal{V}\}$, 
the complete multinomial likelihood has the following form: 
\begin{eqnarray}
l_M^c(\lambda, p, F_1,F_2)&=&k_q\left[m_0\log p+m_+\log(1-p)+n_0\log\{p(1-\lambda)\}+n_+\log\{1-p(1-\lambda)\}\right]\nonumber\\
&&+\sum_{h\in I_q}\sum_{i=1}^{m_+} \left\{m_{i1}(t_h)\log F_{1}(t_h)+ m_{i2}(t_h)\log \bar F_{1}(t_h)\right\}\nonumber\\
&&+\sum_{h\in I_q}\sum_{j=1}^{n_+} (1-V_{jh}) \left[n_{j1}(t_h)\log \{(1-\lambda^*) F_{1}(t_h)\}
+n_{j2}(t_h)\log \{(1-\lambda^*) \bar F_{1}(t_h)\}\right] \nonumber\\
&&+\sum_{h\in I_q}\sum_{j=1}^{n_+}V_{jh} \left[  n_{j1}(t_h)\log \{\lambda^* F_{2}(t_h)\}+n_{j2}(t_h) \log\{\lambda^* \bar F_2(t_h)\}\right].  
\label{cln}
\end{eqnarray}

In the E-step of the $r$th iteration, for $h=1,\ldots,k$ and $j=1,\ldots,n_+$, 
we need to calculate 
\[
Q(\bTheta|\bTheta^{(r-1)})=
E\left\{l_M^c(\lambda, p, F_1,F_2)|\mathcal{X}, \bTheta^{(r-1)}\right\},
\]
where the expectation is with respect to the conditional distribution of $\mathcal{V} $
given $\mathcal{X}$ and substituting  $\bTheta^{(r-1)}$ for $\bTheta$.
With $n_{j1}(t_h)\sim (1-\lambda^*)\text{Bin}\Big(1,F_{1}(t_h)\Big)+ \lambda^*\text{Bin}\Big(1,F_{2}(t_h)\Big)$, it can be checked that 
$$
E(V_{jh}|\mathcal{X},\bTheta^{(r-1)})
=\left\{ a_{h}^{(r)}\right\}^{ n_{j1}(t_h) }\left\{ b_{h}^{(r)}\right\}^{ n_{j2}(t_h)}. 
$$
Therefore,
\begin{eqnarray}
Q(\bTheta|\bTheta^{(r-1)})&=&k_q\left[m_0\log p+m_+\log(1-p)+n_0\log\{p(1-\lambda)\}+n_+\log\{1-p(1-\lambda)\}\right]\nonumber\\
&&+\sum_{h\in I_q}\sum_{i=1}^{m_+} \left\{m_{i1}(t_h)\log F_{1}(t_h)+ m_{i2}(t_h)\log \bar F_{1}(t_h)\right\}\nonumber\\
&&+\sum_{h\in I_q}\sum_{j=1}^{n_+} \left[(1-a_{h}^{(r)})n_{j1}(t_h)\log \{(1-\lambda^*) F_{1}(t_h)\}
+(1-b_{h}^{(r)})n_{j2}(t_h)\log \{(1-\lambda^*) \bar F_{1}(t_h)\}\right] \nonumber\\
&&+\sum_{h\in I_q}\sum_{j=1}^{n_+} \left[ a_{h}^{(r)} n_{j1}(t_h)\log \{\lambda^* F_{2}(t_h)\}+b_{h}^{(r)}n_{j2}(t_h) \log\{\lambda^* \bar F_2(t_h)\}\right].  
\label{cln.qfun}
\end{eqnarray}

In the M-step, 
we update $(\lambda, p, F_1,F_2)$ by 
$$
(\lambda^{(r)},p^{(r)},F_1^{(r)},F_2^{(r)})=\arg\max_{(\lambda, p, F_1,F_2)\in \Theta} Q(\bTheta|\bTheta^{(r-1)}).
$$
After some algebra work, it can be shown that 
$$
Q(\bTheta|\bTheta^{(r-1)})
=Q_1(\lambda)+Q_2(p)+Q_3(F_1)+Q_4(F_2),
$$
where 
\begin{eqnarray*}
Q_1(\lambda)&=&k_q n_0\log (1-\lambda) 
+
\sum_{h\in I_q}\left\{(1-a_{h}^{(r)})n_{1}(t_h)+ (1-b_{h}^{(r)})n_{2}(t_h)\right\} \log(1-\lambda)\\
&&+\sum_{h\in I_q}
\left\{a_{h}^{(r)}n_{1}(t_h)+ b_{h}^{(r)}n_{2}(t_h)\right\} \log \lambda,\\
 Q_2(p)&=&k_q\left[m_0\log p+m_+\log(1-p)+n_0\log p\right]\\
 &&+
 \sum_{h\in I_q}\left\{(1-a_{h}^{(r)})n_{1}(t_h)+ (1-b_{h}^{(r)})n_{2}(t_h)\right\} \log(1-p),\\
Q_3(F_1)&=&  \sum_{h\in I_q}\left[\left\{m_{1}(t_h)+ (1-a_{h}^{(r)})n_{1}(t_h)\right\}\log F_{1}(t_h)
+\left\{m_{2}(t_h)+ (1-b_{h}^{(r)})n_{2}(t_h)\right\}\log \bar F_{1}(t_h)
\right], \\
Q_4(F_2)&=&\sum_{h\in I_q}\left[ a_{h}^{(r)} n_{1}(t_h)\log   F_{2}(t_h) +b_{h}^{(r)}n_{2}(t_h) \log  \bar F_2(t_h) \right]. 
\end{eqnarray*}

Hence, we can update $\lambda$ via  
$$
\lambda^{(r)}=\arg\max_{\lambda} Q_1(\lambda)=\frac{1}{k_qn}\sum_{h\in I_q}\left\{n_1(t_h)a_{h}^{(r)}+n_2(t_h)b_{h}^{(r)}\right\}
$$
and update $p$ via
$$
p^{(r)}=\arg\max_{\lambda} Q_2(p)=\frac{m_0+n_0}{m+n-n\lambda^{(r)}}.
$$
To update $F_1$ and $F_2$, we have 
$$
F_1^{(r)}=\arg\max_{F_1\mbox{ is a cdf}} Q_3(F_1),~~F_2^{(r)}=\arg\max_{F_2\mbox{ is a cdf}} Q_4(F_2).
$$
Following \citet{Dykstra95}, 
we can obtain $F_1^{(r)}$ and $F_2^{(r)}$ via 
\begin{eqnarray*}
F_1^{(r)}&=&\arg\min_{F\mbox{ is a cdf}}\sum_{h\in I_q}
\left\{
m_++n_1(t_h)\left(1-a_h^{(r)}\right)+n_2(t_h)\left(1-b_h^{(r)}\right)
 \right\}
 \left\{
\tilde F_1^{(r)}(t_h)
 -F(t_h)
 \right\}^2,\\
F_2^{(r)}&=&\arg\min_{F\mbox{ is a cdf}}\sum_{h\in I_q}
\left\{
n_1(t_h) a_h^{(r)}+n_2(t_h)b_h^{(r)}
 \right\}
 \left\{
\tilde F_2^{(r)}(t_h)
 -F(t_h)
 \right\}^2,
\end{eqnarray*}
where 
$$
\tilde F_1^{(r)}(t_h)
=
 \frac{m_1(t_h)+n_1(t_h)\{1-a_h^{(r)}\}}
 {m_++n_1(t_h)\{1-a_h^{(r)}\}+n_2(t_h)\{1-b_h^{(r)}\}}
\mbox{
 and 
 }
\tilde F_2^{(r)}(t_h)
=
 \frac{n_1(t_h)a_h^{(r)}}
 {n_1(t_h) a_h^{(r)}+n_2(t_h)b_h^{(r)}}.
 $$

\begin{table}[!p]
\caption{MSEs ($\times1000$) of $\hat{\lambda}$, $\tilde{\lambda}$, $\hat{p}$, and $\tilde{p}$ under Scenario 1.}
\begin{center}
\begin{tabular}{crrrrrrrrr}
\hline
$\lambda$ & 0.25 & 0.25 & 0.25 & 0.5 & 0.5 & 0.5 & 0.75 & 0.75 & 0.75\\
$p$ & 0.25 & 0.5 & 0.75 & 0.25 & 0.5 & 0.75 & 0.25 & 0.5 & 0.75\\
\hline
&\multicolumn{9}{c}{$m=100$, $n=100$}\\
\hline
MSEs of $\hat{\lambda}$      & 18.15 & 9.31 & 4.89 & 12.87 & 7.51 & 4.45 & 7.12 & 4.46 & 2.64\\
MSEs of $\tilde{\lambda}$  & 30.57 & 14.32 & 6.54 & 25.95 & 10.70 & 4.96 & 11.53 & 5.52 & 2.73\\
MSEs of $\hat{p}$                & 1.44 & 2.04 & 1.56 & 1.57 & 2.27 & 1.87 & 1.90 & 2.50 & 1.78\\
MSEs of $\tilde{p}$            & 1.84 & 2.39 & 1.81 & 1.76 & 2.54 & 1.98 & 2.03 & 2.62 & 1.81\\
\hline
&\multicolumn{9}{c}{$m=150$, $n=250$}\\
\hline
MSEs of $\hat{\lambda}$      & 8.55 & 4.06 & 1.99 & 5.95 & 3.07 & 1.80 & 2.82 & 1.56 & 1.15\\
MSEs of $\tilde{\lambda}$  & 18.47 & 8.08 & 2.88 & 13.48 & 4.80 & 2.24 & 5.38 & 2.22 & 1.25\\
MSEs of $\hat{p}$                & 0.92 & 1.27 & 0.97 & 0.96 & 1.37 & 1.07 & 1.18 & 1.54 & 1.05\\
MSEs of $\tilde{p}$            & 1.26 & 1.71 & 1.22 & 1.22 & 1.64 & 1.22 & 1.33 & 1.73 & 1.11\\
\hline
\end{tabular}
\end{center}
\label{table1.malaria}
\end{table}

\begin{table}[!p]
\caption{Kolmogorov--Smirnov  distance ($\times$100) between
the estimated cumulative distribution function
and true cumulative distribution function under
Scenario 1.}
\begin{center}
\begin{tabular}{crrrrrrrrr}
\hline
$\lambda$ & 0.25 & 0.25 & 0.25 & 0.5 & 0.5 & 0.5 & 0.75 & 0.75 & 0.75\\
$p$ & 0.25 & 0.5 & 0.75 & 0.25 & 0.5 & 0.75 & 0.25 & 0.5 & 0.75\\
\hline
&\multicolumn{9}{c}{$m=100$, $n=100$}\\
\hline
$|\hat{F}_1-F_1|_{\infty}$       & 8.36 & 10.18 & 14.35 & 8.72 & 10.55 & 15.10 & 8.87 & 10.98 & 15.65\\
$|\tilde{F}_1-F_1|_{\infty}$ & 9.26 & 11.26 & 15.75 & 9.19 & 11.28 & 15.85 & 9.09 & 11.29 & 16.07\\
$|\hat{F}_2-F_2|_{\infty}$     & 30.06 & 26.57 & 22.45 & 17.30 & 15.27 & 13.29 & 11.41 & 10.58 & 9.61\\
$|\tilde{F}_2-F_2|_{\infty}$ & 47.17 & 34.43 & 25.50 & 23.59 & 17.11 & 13.61 & 12.82 & 10.95 & 9.64\\
\hline
&\multicolumn{9}{c}{$m=150$, $n=250$}\\
\hline
$|\hat{F}_1-F_1|_{\infty}$       & 6.40 & 7.82 & 11.35 & 6.78 & 8.20 & 11.89 & 7.21 & 8.78 & 12.61\\
$|\tilde{F}_1-F_1|_{\infty}$ & 7.67 & 9.42 & 13.57 & 7.57 & 9.23 & 13.30 & 7.64 & 9.26 & 13.45\\
$|\hat{F}_2-F_2|_{\infty}$      & 22.56 & 18.97 & 15.44 & 12.26 & 10.54 & 9.10 & 7.96 & 7.10 & 6.43\\
$|\tilde{F}_2-F_2|_{\infty}$  & 39.82 & 27.90 & 18.34 & 18.07 & 12.34 & 9.60 & 9.48 & 7.53 & 6.48\\
\hline
\end{tabular}
\end{center}
\label{table2.malaria}
\end{table}

\begin{table}[!p]
\caption{MSEs ($\times1000$) of $\hat{\lambda}$, $\tilde{\lambda}$, $\hat{p}$, and $\tilde{p}$ under Scenario 2.}
\begin{center}
\begin{tabular}{crrrrrrrrr}
\hline
$\lambda$ & 0.25 & 0.25 & 0.25 & 0.5 & 0.5 & 0.5 & 0.75 & 0.75 & 0.75\\
$p$ & 0.25 & 0.5 & 0.75 & 0.25 & 0.5 & 0.75 & 0.25 & 0.5 & 0.75\\
\hline
&\multicolumn{9}{c}{$m=100$, $n=100$}\\
\hline
MSEs of $\hat{\lambda}$ & 17.12 & 8.33 & 4.33 & 11.84 & 6.80 & 3.90 & 6.19 & 3.41 & 2.75\\
MSEs of $\tilde{\lambda}$ & 28.35 & 13.98 & 6.39 & 27.10 & 10.39 & 4.74 & 12.76 & 4.87 & 3.05\\
MSEs of $\hat{p}$ & 1.70 & 1.97 & 1.55 & 1.77 & 2.20 & 1.73 & 1.81 & 2.25 & 1.76\\
MSEs of $\tilde{p}$ & 2.12 & 2.43 & 1.89 & 2.02 & 2.52 & 1.92 & 1.98 & 2.45 & 1.86\\
\hline
&\multicolumn{9}{c}{$m=150$, $n=250$}\\
\hline
MSEs of $\hat{\lambda}$ & 11.04 & 4.69 & 2.22 & 6.91 & 3.70 & 1.98 & 3.84 & 2.02 & 1.23\\
MSEs of $\tilde{\lambda}$ & 19.74 & 7.62 & 2.85 & 13.04 & 4.93 & 2.13 & 5.44 & 2.26 & 1.26\\
MSEs of $\hat{p}$ & 0.99 & 1.39 & 1.03 & 1.04 & 1.44 & 1.19 & 1.16 & 1.58 & 1.13\\
MSEs of $\tilde{p}$ & 1.31 & 1.76 & 1.20 & 1.27 & 1.63 & 1.26 & 1.27 & 1.65 & 1.14\\
\hline
\end{tabular}
\end{center}
\label{table3.malaria}
\end{table}

\begin{table}[!p]
\caption{Kolmogorov--Smirnov  distance ($\times$100) between
the estimated cumulative distribution function
and true cumulative distribution function under
Scenario 2.}
\begin{center}
\begin{tabular}{crrrrrrrrr}
\hline
$\lambda$ & 0.25 & 0.25 & 0.25 & 0.5 & 0.5 & 0.5 & 0.75 & 0.75 & 0.75\\
$p$ & 0.25 & 0.5 & 0.75 & 0.25 & 0.5 & 0.75 & 0.25 & 0.5 & 0.75\\
\hline
&\multicolumn{9}{c}{$m=100$, $n=100$}\\
\hline
$|\hat{F}_1-F_1|_{\infty}$      & 8.29 & 9.96 & 14.90 & 8.55 & 10.71 & 15.51 & 8.97 & 11.14 & 15.72\\
$|\tilde{F}_1-F_1|_{\infty}$  & 9.34 & 11.27 & 16.44 & 9.23 & 11.53 & 16.58 & 9.20 & 11.58 & 16.32\\
$|\hat{F}_2-F_2|_{\infty}$     & 30.38 & 26.27 & 22.77 & 17.27 & 15.53 & 13.38 & 11.86 & 10.82 & 9.91\\
$|\tilde{F}_2-F_2|_{\infty}$ & 47.76 & 35.96 & 26.90 & 25.30 & 18.11 & 13.97 & 14.54 & 11.50 & 10.02\\
\hline
&\multicolumn{9}{c}{$m=150$, $n=250$}\\
\hline
$|\hat{F}_1-F_1|_{\infty}$      & 6.51 & 8.06 & 11.49 & 6.99 & 8.56 & 11.94 & 7.22 & 8.89 & 12.73\\
$|\tilde{F}_1-F_1|_{\infty}$  & 7.49 & 9.23 & 13.02 & 7.55 & 9.23 & 12.90 & 7.45 & 9.19 & 13.12\\
$|\hat{F}_2-F_2|_{\infty}$      & 23.72 & 19.89 & 15.90 & 12.64 & 10.61 & 8.76 & 7.78 & 6.77 & 6.28\\
$|\tilde{F}_2-F_2|_{\infty}$  & 37.46 & 25.50 & 17.52 & 16.55 & 11.65 & 8.93 & 8.51 & 6.90 & 6.29\\
\hline
\end{tabular}
\end{center}
\label{table4.malaria}
\end{table}

\begin{table}[!p]
\caption{\label{example.table} Point estimates along with bootstrap SEs of $\lambda$ and $p$ and 95\% BPCIs of 
$\lambda$ and $p$ for the malaria data.
}
\begin{center}
\begin{tabular}{ccccc}
\hline
&\multicolumn{2}{c}{$\lambda$}&\multicolumn{2}{c}{$p$}\\
&Our method&Binomial estimator&Our method&Binomial estimator\\
\hline
Point estimate &0.545&0.541&0.439&0.438\\  
SE of point estimate & 0.063&0.0747&0.039&0.042\\ 
95\% BPCI &$[0.424,0.670]$&$[0.382,0.669]$&$[0.364,0.520]$&$[0.347,0.514]$\\  
\hline\end{tabular}
\end{center}
\end{table}

\begin{figure}[!p]
\centering
\includegraphics[scale=0.8]{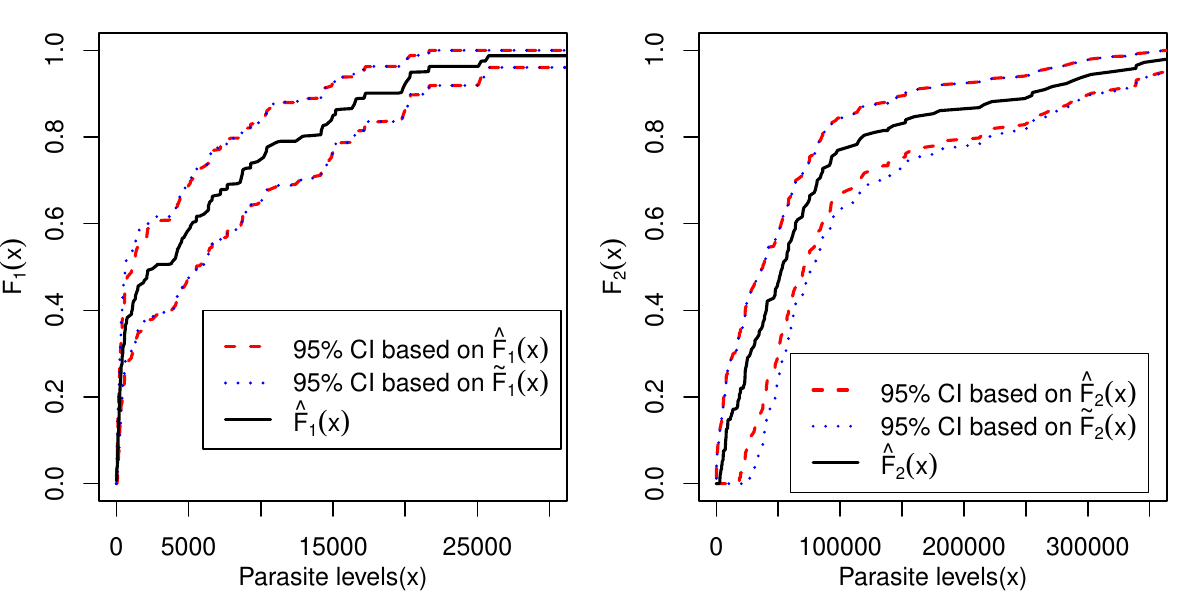}
\caption{95\% bootstrap pointwise confidence bands of $ {F}_1$ and $F_2$ based on our method and the plug-in method.}
\label{cdf.malaria}
\end{figure}

\begin{figure}[!p]
\centering
\includegraphics[scale=0.5]{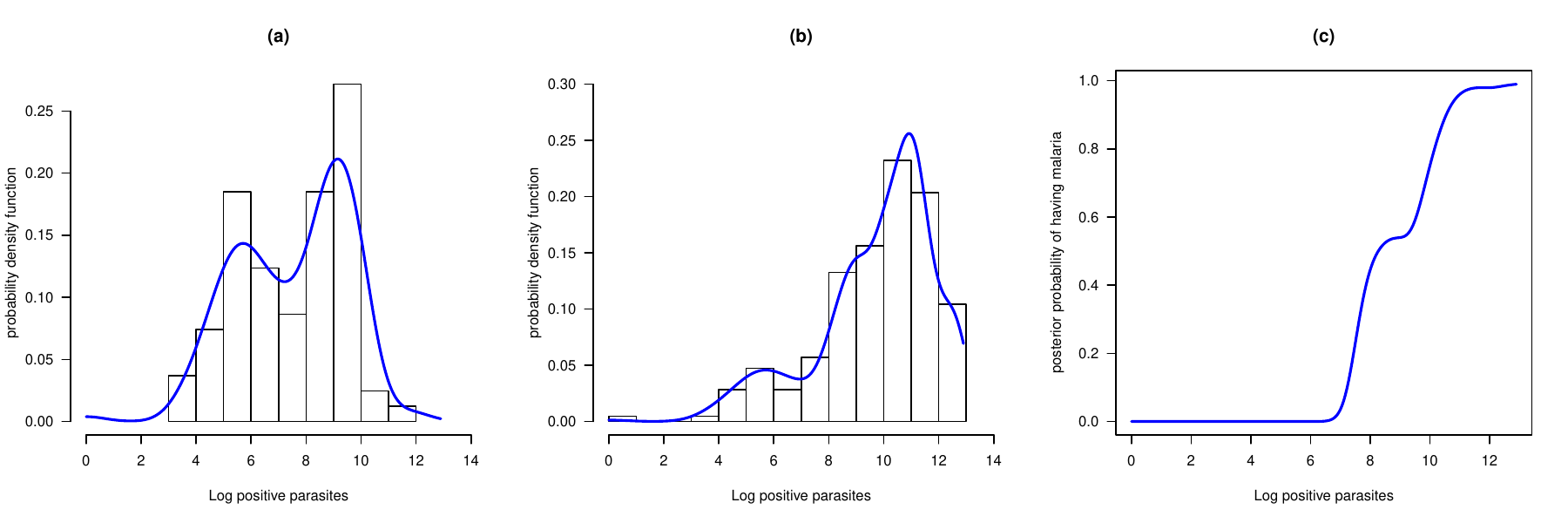}
\caption{Density and posterior probability estimation of the malaria data:
Panel (a) plots the histogram of the logarithm of the positive parasite levels in the nonmalaria population and the density estimate $\hat g_1$; 
panel (b) plots the histogram of the logarithm of the positive parasite levels in the mixture sample 
and the density estimate $(1-\hat\lambda^*)\hat g_1+\hat \lambda^*\hat g_2$; 
and panel (c) plots the estimated posterior probability of catching malaria given the logarithm of the positive parasite level in the mixture sample. }
\label{den.malaria}
\end{figure}

\end{document}